# Superconducting tunable flux qubit with direct readout scheme

Fabio Chiarello[1], Pasquale Carelli[2], Maria Gabriella Castellano[1], Carlo Cosmelli[3], Lorenzo Gangemi[3], Roberto Leoni[1], Stefano Poletto[1,4], Daniela Simeone[1,4], Guido Torrioli[1]

[1] Istituto di Fotonica e Nanotecnologie, CNR, via Cineto Romano 42, I-00156 Roma, Italy

[2] Dipartimento di Energetica, Università dell'Aquila, Monteluco di Roio, I-67040 L'Aquila, Italy

[3] Dipartimento di Fisica, Università di Roma "La Sapienza," P.le Aldo Moro 5, I-00185 Roma, Italy

[4] Dipartimento di Fisica, Università di Roma 3, via della Vasca navale 84, I-00146 Roma, Italy

E-mail: *chiarello@ifn.cnr.it*

We describe a simple and efficient scheme for the readout of a tunable flux qubit, and present preliminary experimental tests for the preparation, manipulation and final readout of the qubit state, performed in incoherent regime at liquid Helium temperature. The tunable flux qubit is realized by a double SQUID with an extra Josephson junction inserted in the large superconducting loop, and the readout is performed by applying a current ramp to the junction and recording the value for which there is a voltage response, depending on the qubit state. This preliminary work indicates the feasibility and efficiency of the scheme.

*24 June 2005*

## 1. INTRODUCTION

Quantum computation is a novel and promising architecture that overcomes the intrinsic limitations of the classical one [1]. The quantum bit (or qubit), basic element of quantum computers, is a two state quantum system that can be prepared in a defined initial state, coherently manipulated by unitary transformations, and finally measured. Qubits can be implemented by using very different physical systems. In particular solid-state devices are promising for the large-scale integration and for the individual control and readout of many qubits. Single [2-7] and entangled couples [8,9] of superconducting solid-state qubits based on the Josephson effect have been realized and studied. In some of these systems the integration of the readout in the qubit [4-7] improves the system compactness, simplicity and efficiency. In this direction, we present a Josephson qubit based on a double SQUID device with a modification allowing the integrated, direct readout of its magnetic flux state.

## 2. TUNABLE FLUX QUBIT

A double SQUID (Superconducting QUantum Interference Device) consists of a superconducting loop of total inductance $L$ interrupted by a small dc-SQUID, formed by a second superconducting loop of inductance $\ell$ interrupted by two identical Josephson junctions, each with critical current $i_0$ and capacitance $c$ [10]. The device can be biased by two magnetic fluxes, $\Phi_x$ applied to the large loop and $\Phi_c$ applied to the dc-SQUID respectively (fig. 1a). If the dc-SQUID loop is small enough (for $\ell \ll i_0 \Phi_b$, where $\Phi_b = \Phi_0/2\pi$ and $\Phi_0 \cong 2.068 \times 10^{-12}\, Wb$ is the flux quantum), the inner dc-SQUID behaves approximately like a single junction with tunable critical current $I_0 = 2i_0 |\cos(\pi \Phi_c / \Phi_0)|$ and capacitance $C = 2c$, so that the double SQUID can be approximately replaced by a simple rf-SQUID with tunable critical current. The dynamics is described by the phase difference $\varphi$ across the dc-SQUID, related to the total magnetic flux threading the large loop $\Phi = \Phi_b \varphi$ and to the current circulating in it, $I_q = -(\varphi - \varphi_x)\Phi_b / L$ (where $\varphi_x = \Phi_x / \Phi_b$ is the reduced flux bias). The Hamiltonian $H = T + U$ is the sum of the kinetic term $T = Q^2/2C$ (Q is the total charge on the junction capacitance) and of the potential:

$$U = \frac{\Phi_b^2}{2L}(\varphi - \varphi_x)^2 - I_0 \Phi_b \cos(\varphi) \qquad (1)$$

For $j_x = p$ (corresponding to $\Phi_x = \Phi_0/2$) and $b_L = LI_0/\Phi_b > 1$ the potential is symmetric, with two identical wells separated by a central barrier whose height is determined by $b_L$. The potential shape is modified by the applied fluxes: $\Phi_x$ controls the symmetry of the potential (fig. 1b), while $\Phi_c$, modifying the critical current $I_0$ and hence the effective value of $b_L$, modifies the barrier height (fig. 1c).

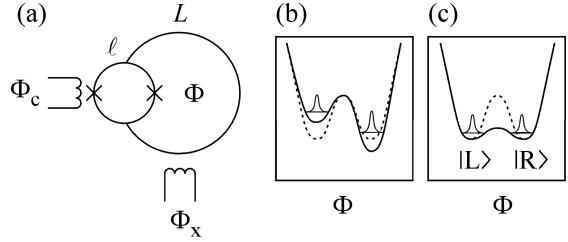

Fig. 1 Scheme of the double SQUID qubit. (b) The symmetric qubit potential (dashed curve) can be tilted (continuous curve) by applying the flux $\Phi_x$. (c) The barrier height can be raised or lowered by applying the flux $\Phi_c$.

This device can be used as a qubit: the computational states are mapped in the two distinct magnetic flux states localized in the left and right minima of the potential ($|L\rangle$ and $|R\rangle$ respectively), corresponding to two different values of the current circulating in the large loop, $I_q^L$ and $I_q^R$. In this basis the Hamiltonian can be written by using the Pauli's matrices $s_x$ and $s_z$:

$$H = -\frac{1}{2}\hbar e(\Phi_x) s_x - \frac{1}{2}\hbar \Delta(\Phi_c) s_z \qquad (2)$$

Where $\hbar e$ is the energy difference between the two minima, controlled by $\Phi_x$, and $\hbar\Delta$ is the spacing between the fundamental and the first excited energy levels, controlled by $\Phi_c$.

The manipulation can be performed either by microwave pulses used to excite the upper state, or by fast variations of the bias fluxes. In this paper we concentrate our attention on this second method, but without excluding the use of microwaves or of a hybrid technique [11-13].

The first step is to prepare the qubit in a determined flux state, i.e. in one specific well of the potential. Without this step, we would find the wavefunction in either state at random. Preparation is performed by unbalancing the potential till it shows only one absolute minimum where the system relaxes (a reduction of the barrier during this phase improves the process), then restoring the balancing. Coherent rotations between the two states can be performed by reducing the barrier height and allowing the state to evolve freely for fractions of the oscillation period, while maintaining the potential symmetry. The rotation of the relative phase between the two states can be achieved by slightly moving the system away from the symmetry while maintaining the barrier high. In this way the full control of the qubit is possible. In principle this control can be efficiently achieved on-chip by using Rapid Single Flux Quantum logic [14,15], which is fast, compact and easily integrable with the quantum elements.

## 3. INTEGRATED DIRECT READOUT

The qubit readout can be done by an inductively coupled magnetometer, for example a shunted or unshunted dc-SQUID [16,17]. However, a simpler and more efficient technique can be implemented by modifying the qubit design interrupting the large loop of the rf-SQUID with a large junction of critical current $I_{0L} \gg 2i_0$, with two terminals for the direct injection of current (fig. 2a) [18]. The design is similar to the "Quantronium" readout scheme [4], but with very different operating principles.

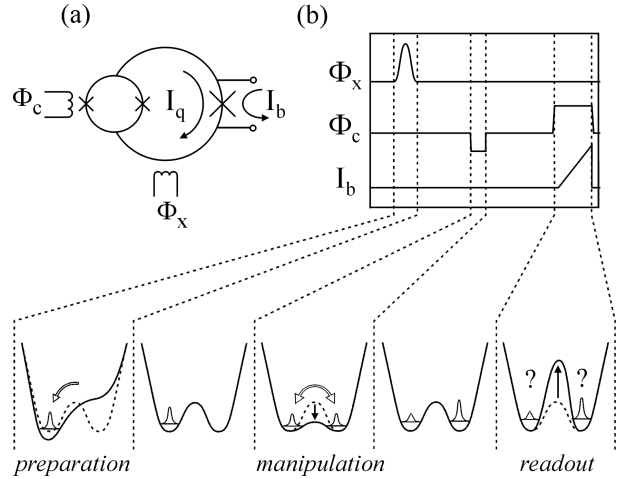

Fig. 2 (a) Direct readout scheme based on the double SQUID qubit with the insertion of a large junction. (b) Timing scheme of bias controls (for a single cycle), showing preparation, manipulation, readout, and frozen intermediate states.

The full Hamiltonian of this system, if the dc-SQUID can be approximated by a controllable junction as described, is given by $H = T + U$, with kinetic contribution $T = Q^2/2C + Q_L^2/2C_L$ ($C_L$ and $Q_L$ are respectively the large junction capacitances and the total charge on it) and potential U:

$$U = \frac{\Phi_b^2}{2L}(\varphi - \delta - \varphi_x)^2 - I_0 \Phi_b \cos\varphi - I_{0L}\Phi_b \cos\delta - I_b \Phi_b \delta \qquad (3)$$

where δ is the phase across the large junction, so that the voltage is $V_L = \Phi_b \, d\delta/dt$.

In the absence of directly injected current $I_b$, the large junction can be neglected (in the limit $I_{0L} \gg 2i_0$). If a current $I_b$ is injected, the junction is crossed by the sum of this current plus the current $I_q$ circulating in the qubit and related to its state ($I_q^L$ in the left state and $I_q^R$ in the right one). For $I_b + I_q < I_{0L}$ the junction remains in the superconducting state, and the qubit behaves like a simple rf-SQUID but with an extra phase bias $\delta\varphi_x \cong \arcsin\left[(I_b + I_q)/I_{0L}\right]$ that causes an unbalance in the potential symmetry. Indeed, a characteristic of this system is the trade off between high sensitivity and invasive readout. As soon as the total current overcomes the critical value, for $I_b + I_q > I_{0L}$, the junction goes to the running state and a voltage develops across its terminals. In a real case, because of thermal and quantum fluctuations, the transition is no more deterministic but it is randomly distributed just below $I_{0L}$, with a mean value $\bar{I}_{0L}$. The qubit state can be read by applying a current ramp to the large junction (from zero to a maximum above the critical current) and recording the value $I_b^*$ corresponding to the voltage transition. This is repeated many times (from 100 to 5000, according to the desired precision) in order to evaluate the mean value $\bar{I}_b^*$ and hence to estimate the qubit current $I_q \cong \bar{I}_{0L} - \bar{I}_b^*$. In order to ensure that the qubit readout is not affected by the extra phase bias $\delta\varphi_x$, it is required that the distinct flux states remain well separated also in the presence of the extra unbalancing; this can be obtained by maintaining the barrier high enough throughout the readout process. The voltage transition occurs when $I_b + I_q \sim I_{0L}$, corresponding to an extra phase $\delta\varphi_x \sim \pi/2$, and the requirement to have separated wells also in this case gives the limit $\beta_L > 2.79$.

## 4. STATE PREPARATION AND MANIPULATION

We have designed microchips containing the double SQUID with integrated readout. Two sets have been realized, one by Hypres Inc. and one by our home facility respectively, using a Nb trilayer process with 100 A/cm$^2$ critical current density. The device is designed in a full gradiometric configuration, with both the large loop and the small dc-SQUID loop gradiometric along orthogonal directions, in order to strongly reduce the flux noise pick-up and the spurious couplings between loops and bias coils. The target parameters of the device are: $L = 85\,pH$, $\ell = 6\,pH$, $C = 0.6\,pF$, $C_L = 5\,pF$, $2i_0 = 15\,\mu A$, $I_{0L} = 100\,\mu A$. The aim of this preliminary work is to characterize the device and to test the techniques for the preparation of the qubit state, the manipulation by means of flux pulses, and its readout. For this purposes it is sufficient to work in the incoherent regime at liquid Helium temperature, a simpler and more controllable condition with respect to the final goal, i.e. in coherent regime at lower temperature.

The qubit is controlled by applying a synchronized sequence of signals for the bias current and the control fluxes, with a repetition period of *50ms*. The sequence allows one to implement initial preparation, intermediate manipulation and final readout; a typical sequence is shown in fig. 2b. The flux $\Phi_x$, controlling the potential symmetry, is maintained to a fixed value $\Phi_x^{base}$ during all the cycle except for a short Gaussian pulse at the beginning: this pulse, whose duration is of the order of *2ms*, has an amplitude chosen to tilt the potential towards the right (left) in such a way to have only one well available where the system can relax. The pulse can be positive or negative, in order to prepare the qubit in either one of the two possible flux states. At the end, the symmetry of the potential is restored to the initial value and the state is in the right (left) well, achieving the desired preparation.

The readout current $I_b$ is maintained to zero during all the cycle except for the last *10ms*, when it is ramped from zero to a value just above the maximum critical current, and then quickly returned to zero. In correspondence of the junction jump to the voltage state, the value $\bar{I}_b^*$ is acquired, allowing one to get the value of $I_q$. During the readout, the flux $\Phi_c$ that controls the barrier height is changed in order to have the maximum barrier, to prevent the destruction of the qubit state by the extra phase bias. During the rest of the cycle, the flux $\Phi_c$ is kept to a base value $\Phi_c^{base}$ just sufficient to ensure a good separation between the two states, except halfway between preparation and readout: here the barrier is decreased for a short time $\Delta t$ (chosen between *20ns* and *2ms*) in order to make possible free evolution between the two flux states.

The readout allows the efficient one-shot discrimination of the qubit state, and repeated measurements allow estimating the probability of obtaining one of the two states after the manipulation.

In a first test we checked the qubit preparation and readout, with no manipulating pulse for $\Phi_c$ in the middle. The base value of the control $\Phi_c^{base}$ is fixed, and repeated series of preparation/readout cycles are performed for different values of the base flux $\Phi_x^{base}$. The continuous curve of fig. 3 shows the qubit current $I_q$ versus the applied flux $\Phi_x^{base}$ after the system has been prepared in one of the two flux states (say the left state), the dashed curve shows the same curve for a preparation in the other state (right state). For both curves, the value of $\Phi_c^{base}$ is the one corresponding to the highest possible barrier for that particular experimental sample. The shapes of these curves allow a rough estimation of $b_L^{Max} \approx 3.9$ for the maximum barrier. The curves are periodic with period equal to a single flux quantum $\Phi_0$. In a single period it is possible to distinguish three zones. In the first zone, indicated in fig. 3 with (a), the potential has just one minimum, so that the two measured characteristics coincide. In regions (b) and (c) there are two distinct minima that can be occupied or not according to the initial preparation. In region (c) this is shown by the two distinct characteristics, each corresponding to one of the two possible states. In region (b), despite the existence of two distinct states, the two characteristics coincide because of the measurement effect. In fact in this region the extra phase introduced by the measurement is sufficient to unbalance the potential so that an initial left state is turned into a right state (while the right state is not changed), so that the final result is always the right one. The dotted curve in fig. 3 is a reconstruction of the true characteristics in the absence of this spurious effect. It is evident that the symmetric position, which corresponds to a symmetric double well potential, is well inside the (c) zone, allowing the correct qubit operations.

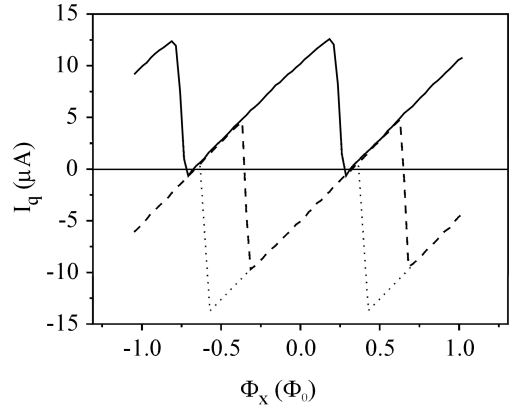

Fig. 3 Measured characteristic of the direct readout flux qubit in the absence of intermediate manipulating pulses, with initial preparation in the left state (continuous curve), and in the opposite state (dashed curve). It is enlightened the missing part due to the measurement procedure (dotted curve).

In a second test, we introduced the intermediate manipulating pulse in $\Phi_c$ that reduces the barrier height for a fixed time $\Delta t = 2 ms$ and allows oscillations between the two states. The base value of the unbalancing flux $\Phi_x^{base}$ is chosen such as to have the potential as symmetric as possible during the manipulation, compatibly with the experimental limits. The observed $I_q$ can assume two possible distinct values, corresponding to the two possible qubit states.

In fig. 4 it is plotted the probability $P$ to observe the system in the right state, once prepared in the right (dashed curve) or left state (continuous curve), in function of the barrier height (expressed in term of $b_L/b_L^{Max}$, equal to $|\cos(p\,\Phi_c/\Phi_0)|$). For high barrier (higher $b_L/b_L^{Max}$) the system remains in the prepared state (either right or left), while for small barrier (lower $b_L/b_L^{Max}$) the system evolves to equilibrium. In this case, the equilibrium value is not exactly 50% due to a residual asymmetry of the potential during the system evolution. The continuous curve corresponds to an initial preparation in the left state, the dashed one to an initial preparation in the right state.

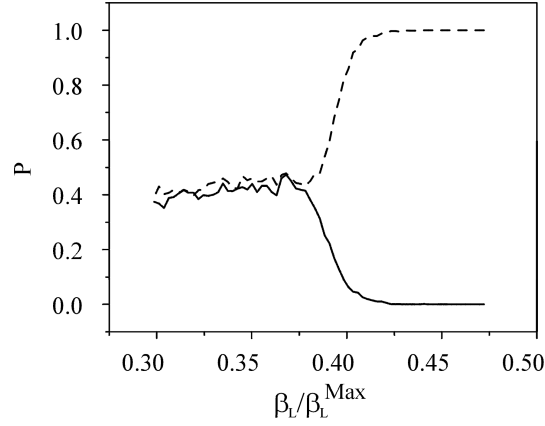

Fig. 4 Measured probability to find the system in the right state for different barrier heights, expressed in terms of $b_L / b_L^{Max}$.

The third measurement consists in acquiring the probability $P$ to obtain the right state after a left preparation for different duration of the manipulating pulse $\Delta t$, using the previous set-up for different fixed $b_L / b_L^{Max}$. This allows following the incoherent evolution of the qubit population (fig. 5).

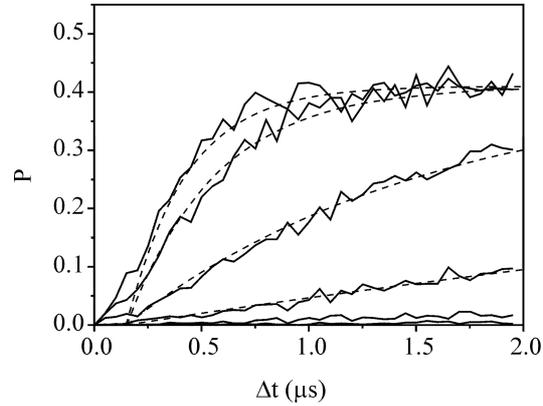

Fig. 5 Measured evolution of the probability to obtain a right state after a preparation in the left state, as a function of the manipulating pulse duration $\Delta t$ (straight curves), and fit with exponential relaxations (dashed curves), for $b_L / b_L^{Max}$ values 0.35, 0.36, 0.37, 0.39, 0.40, 0.41 (from the uppermost to the lowest curve).

In the chosen test regime we observe incoherent relaxations (fig. 5) with characteristic times that can be estimated by fitting the curves with exponential relaxations obtaining *0.2μs*, *0.42μs*, *1.4μs* and *7μs* for $b_L / b_L^{Max}$ equal to 0.35, 0.36, 0.37, 0.39, 0.40 and 0.41 respectively (from the uppermost to the lowest curves in fig. 5). In principle this technique should allow the observation of coherent oscillations in the future stage, in quantum regime.

## 5. CONCLUSIONS

The working principles of a tunable flux qubit with a direct readout junction have been tested, showing the feasibility and the effectiveness of the device, and allowing to set up the apparatus towards the test at low temperature for the study in the quantum regime.

This work is supported by the European Community Project RSFQubit, by the INFN SQC Project and by the MIUR-FIRB project "Nanotechnologies and Nanodevices for the information society".